\documentstyle[osa,manuscript,psfig]{revtex}

\begin{document}

\title{Spin alignment of vector mesons in 
unpolarized hadron-hadron collisions at high energies}

\author {Xu Qing-hua, and Liang Zuo-tang}
\address{Department of Physics,
Shandong University, Jinan, Shandong 250100, China \\}

\maketitle

\begin{abstract}
We argue that spin alignment of the vector mesons observed in unpolarized
hadron-hadron collisions is closely related to the single spin 
left-right asymmetry observed in transversely 
polarized hadron-hadron collisions.
We present the numerical results 
obtained from the type of spin-correlation imposed by the 
existence of the single-spin left-right asymmetries.
We compare the results with the available data and make predictions
for future experiments.
\end{abstract}


\newpage

\section{Introduction}

It has been observed\cite{pH} for a long time that
hyperons with moderately large transverse momenta
in unpolarized hadron-hadron collisions are polarized transversely
to the production plane.
Data for different hyperons
at different energies in reactions using different projectiles and/or
targets have been available\cite{pHdata}.
Different efforts\cite{Theor,LB97,ANS01} have been made to understand them. 
Progress has been achieved in last years, but
the role of spin in the production dynamics is still not well understood.
It should be interesting and helpful to see whether similar effects
exist for the production of other kinds of hadrons.
In this connection, we note 
the measurements on the polarization of vector mesons 
along the normal of the production plane
in unpolarized hadron-hadron 
or hadron-nucleus collisions\cite{Savhh,BCGMNS,EXCHARM00}, 
especially those of $K^{*+}$ in $K^+p$ 
and neutron-carbon collisions\cite{BCGMNS,EXCHARM00}.
The obtained results show
an obvious ``spin alignment'' for $K^{*+}$. 
Clearly, the study of them may shed new light on the searching
of the origin of the hyperon polarization in unpolarized
hadron-hadron collisions.

In this note, we study the spin alignment of vector mesons
in unpolarized hadron-hadron collisions by relating it to 
the single spin left-right asymmetries ($A_N$) observed in 
transversely polarized hadron-hadron collisions\cite{ANdata}.
We argue that these two effects are closely related to each
other and make calculations for the spin alignment of vector mesons
using the same method as that used 
in $e^+e^-$ annihilation at $Z^0$ pole\cite{XLL01}. 
We compare the results obtained with 
the available data\cite{Savhh,BCGMNS,EXCHARM00}  
and make predictions for future experiments.

\section{Qualitative arguments}

\subsection{Spin alignment of a vector meson 
in the fragmentation of a polarized quark}
 
The polarization of a vector meson is 
described by the spin density matrix $\rho$ 
or its element $\rho_{mm'}$,
where $m$ and $m'$ label the spin component along 
the quantization axis.
The diagonal elements $\rho_{11}$, $\rho_{00}$ and $\rho_{-1-1}$
for the unit-trace matrix $\rho$ are the relative intensities for $m$ 
to take the values $1$, $0$, and $-1$ respectively.
In experiment, $\rho_{00}$ can be measured from the angular
distribution of the decay products of the vector mesons. 
If the meson is unpolarized, we have 
$\rho_{00}=\rho_{11}=\rho_{-1-1}=1/3$. 
Hence, a deviation of $\rho_{00}$ from $1/3$ indicates 
that the spin of the vector meson has a larger 
(if $\rho_{00}<1/3$, or smaller if  $\rho_{00}>1/3$) 
probability to be parallel of anti-parallel to 
the quantization direction. 
Such a phenomenon is referred as spin alignment 
of the vector meson along the quantization axis. 

Recently, high accuracy data on $\rho_{00}$ have been published for
different vector mesons 
in $e^+e^-$ annihilation at $Z^0$ pole\cite{Kress01}.
The results show a clear spin
alignment for the vector mesons 
along the moving direction of the produced quark
in particular in the large momentum fraction region.
The $\rho_{00}$ measured in the helicity frame is, e.g., larger
than 0.5 for $z>0.5$, where $z\equiv2p_V/\sqrt{s}$, $p_V$ is the
momentum of the vector meson, 
$\sqrt{s}$ is the total $e^+e^-$ center of mass (c.m.) energy.
We note that the vector mesons with large $z$ 
are predominately fragmentation
results of the initial quarks, and 
according to the Standard Model for electroweak interactions
the initial quarks produced at $e^+e^-$ annihilation
vertex at $Z^0$ pole are longitudinally polarized.\cite{Augstin}
These data\cite{Kress01} lead us to the following conclusion:
In the longitudinally polarized case, 
the vector mesons produced
in the fragmentation of a polarized quark have a significant spin
alignment with the fragmenting quark.
The resulting $\rho_{00}$ for the vector mesons is significantly
larger than $1/3$ in the frame where the polarization direction
of the fragmenting quark is chosen as the quantization axis.

The above-mentioned conclusion
is derived from the $e^+e^-$ annihilation data\cite{Kress01}, 
and applies to the fragmentation of a longitudinally polarized quark. 
If we now extend it to  
the fragmentation of a quark polarized in any direction, 
we obtain the following general conclusion: 
{\it In the fragmentation of a polarized quark,
the vector mesons produced
have a significant spin alignment with the fragmenting quark.
The resulting $\rho_{00}$ for the vector meson is significantly
larger than $1/3$ in the frame where the polarization direction
of the fragmenting quark is chosen as the quantization axis.}

We should emphasize that the extension of the above-mentioned 
conclusion from the longitudinally polarized case to the 
transversely polarized case is not obvious. 
This can be seen from the difference between the 
helicity distribution of the quarks in longitudinally 
polarized nucleon and the transversity distribution\cite{BDR2002}. 
Because of the relativistic effects, the magnitudes 
and/or shapes 
(i.e. the $x$-dependences, where $x$ is the fractional 
momentum carried by the quark in the nucleon)
of them are in general different from each other. 
But it seems that the qualitative features in particular 
the signs of them are the same, especially in 
the large $x$ region\cite{BDR2002,MA1998}.  
The relativistic effects do not change the signs of them.
Similarly, we expect that the magnitude and the $z$-dependence 
of $\rho_{00}$ 
for the vector meson obtained 
in the fragmentation of a longitudinally polarized 
quark can be different from those in the fragmentation 
of a transversely polarized quark, even if the quantization 
axis is chosen as the polarization direction in each case. 
(Here, $z$ is the fractional momentum carried by the vector meson.)
But, we also expect that the qualitative features 
in particular whether $\rho_{00}$ is larger or smaller than $1/3$
should be the same.  
The differences in the two cases should be reflected in 
the calculations of $\rho_{00}$ as a function of $z$ 
and we will come back to this point in next section when 
formulating the calculation method. 
Since we are, at the present stage, more interested in 
the qualitative features, we will assume the conclusion 
is in general true and use it as a starting point 
for the following discussions. 
 
\subsection{Spin alignment of the vector meson 
in unpolarized and left-right asymmetry 
in singly polarized hadron-hadron collisions} 

Having the conclusion obtained above in mind,  
we now ask what we can obtain from the 
data\cite{BCGMNS,EXCHARM00} mentioned in the introduction 
for the spin alignment of $K^{*+}$ 
in unpolarized $K^+p$ 
and neutron-carbon collisions.  
These experiments show in particular that\cite{EXCHARM00}, 
$\rho_{00}=0.424\pm0.011\pm0.018$
for $K^{*+}$ with $p_V>12$GeV 
produced in unpolarized neutron-carbon collisions 
with $p_{inc}=60$GeV.  
We see that the $\rho_{00}$ obtained for 
the vector meson $K^{*+}$ 
is significantly larger than $1/3$ if the normal 
of the production plane is chosen as the quantization axis.
Together with the conclusion obtained above, 
this data leads immediately to the 
following statement: 
There should be a significant polarization 
of the quark which contributes to the
vector meson production and the polarization is transverse 
to the production plane. 
In other words, there exists a spin-correlation of the type
$\vec s_q \cdot \vec n$ in the reaction.
[Here, $\vec s_q$ is the spin of the fragmenting quark;
$\vec n\equiv (\vec p_{inc}\times \vec p_V)/|\vec p_{inc}\times \vec p_V|$ 
is the unit vector in the normal direction of the production plane, 
$\vec p_{inc}$ and $\vec p_V$ are respectively 
the momentum of the incident
hadron and that of the produced hadron.] 
The spin alignment of the vector meson in unpolarized 
hadron-hadron or hadron-nucleus collisions is 
one of the manifestations of the $\vec s_q \cdot \vec n$ type of 
spin correlation in the reaction.

Now, we recall that, in another class of hadron-hadron collision
experiments where a transversely polarized beam is used, 
a significant single-spin left-right asymmetry $A_N$ has been 
observed\cite{ANdata}.
The existence of the non-zero $A_N$ implies a significant spin
correlation of the form $\vec s_P \cdot \vec n$ in such processes,  
where $\vec s_P$ is the spin of the projectile. 
We recall that this effect exists mainly in the fragmentation region
and hadrons in this region are predominately the fragmentation results
of the valence quarks of the incident hadron.
Having in mind that the valence quarks are in general also
transversely polarized if the beam hadron is transversely polarized,
we reach the conclusion that the kind of spin-correlation indicated
by the existence of $A_N$ is very similar to that inferred from
the existence of spin alignment of vector mesons in unpolarized
hadron-hadron collisions.
Hence, we suggest that 
{\it they are two different manifestations of the same 
spin-correlation of the form $\vec s_q \cdot \vec n$ 
in hadron-hadron or hadron-nucleus collisions.} 
This is similar to [\ref{LB97}] where it is suggested that 
hyperon polarization in unpolarized hadron-hadron collisions 
has the same origin as the single-spin 
left-right asymmetry $A_N$. 
Both of them are also manifestations of the 
$\vec s_q\cdot \vec n$ type of spin correlation 
in hadron-hadron or hadron-nucleus collisions.

\section{A quantitative calculation}

Since the purpose of this paper is to 
demonstrate the close relation between the existence of 
$A_N$ and that of spin alignment in unpolarized hadron-hadron collisions 
in an explicit and possibly quantitative manner, 
we follow the same way as that adopted in [\ref{LB97}] 
and do the following calculations. 
We use the results derived from the $A_N$ data\cite{ANdata} 
for the strength of the above-mentioned $\vec s_q\cdot \vec n$ 
type of spin-correlation as input to 
calculate the polarization of the valence-quark 
before the hadronization. 
Then, to calculate $\rho^V_{00}$ 
in unpolarized hadron-hadron collisions, 
we need only to calculate $\rho^V_{00}$ in the hadronization of 
a transversely polarized quark. 
We do the calculation using the same method 
as that used in calculating 
$\rho^V_{00}$ in $e^+e^-$ annihilation at $Z^0$ pole 
where $\rho^V_{00}$ is calculated in the hadronization of 
a longitudinally quark.
The details of each step is given in the following.

\subsection{ Calculating $\rho_{00}$ in $e^+e^-\to Z^0\to VX$}

The method for calculating 
$\rho^V_{00}$ in $e^+e^-$ annihilations at $Z^0$ pole is given 
in Ref.[\ref{XLL01}] and can be summarized as follows:
To do the calculations,
we need to divide the final state vector mesons 
into the following two groups and consider them separately:
(A) those which are directly produced and contain the polarized
fragmenting quark;
(B) the rest.
Then, $\rho_{00}^V$ is given by\cite{XLL01},
\begin{equation}
\rho^V_{00}(z)=
\frac{\rho^V_{00}(A)N_V(z,A)+\rho^V_{00}(B)N_V(z,B)}
     {N_V(z,A)+N_V(z,B)},
\end{equation}
where $N_V(z,A)$ and $N_V(z,B)$ are the number densities of vector mesons
of group (A) and (B) respectively, 
$\rho^V_{00}(A)$ and $\rho^V_{00}(B)$ are their 
corresponding $\rho_{00}^V$'s, 
and $z$ is the momentum fraction carried by the vector meson.      

The vector mesons in group (B) do not contain 
the polarized fragmenting quarks 
and there are in general many different 
possibilities to produce them in a collision process. 
Hence, they are simply taken as unpolarized\cite{XLL01}, 
i.e. $\rho^V_{00}(B)=1/3$. 
For those in group (A), the spin density matrix
$\rho^V(A)$ is obtained\cite{XLL01} from the
direct product of the spin density matrix of 
the fragmenting quark, $\rho^q$, and that 
of the anti-quark, $\rho^{\bar{q}}$, 
which combines with the fragmenting quark 
to form the vector meson. 
Taking the most general form for $\rho^{\bar{q}}$, 
we obtained that\cite{XLL01},
if the polarization direction of the fragmenting quark 
is chosen as the quantization axis,
$\rho^V_{00}(A)$ is given by,
\begin{equation}
\rho^V_{00}(A)=(1-P_{q}P_z)/(3+P_{q}P_z),
\label{eq1}  
\end{equation}
where $P_q$ is the polarization of the fragmenting quark, 
and the polarization direction is taken as $z$-direction; 
$P_z$ is the $z$-component of the polarization of the anti-quark. 
By comparing the results with the 
$e^+e^-$ annihilation data\cite{Kress01},
we see that\cite{XLL01}, to fit the data, we have to
take the anti-quark as polarized in the opposite direction as the
fragmenting quark, and the polarization $P_z$ is proportional
to $P_q$, i.e.,
\begin{equation}
P_z = -\alpha P_q,
\label{eq11}
\end{equation} 
where $\alpha \approx 0.5$ is a constant.
Inserting Eq.(3) into Eq.(2), we obtain that,
$\rho^V_{00}(A)=1/3+\Delta \rho^V_{00}(A)$, and
\begin{equation}
\Delta \rho^V_{00}(A)=4\alpha P_{q}^2/[3(3-\alpha P_{q}^2)].
\label{eqdelt}
\end{equation} 
Using this, we can fit\cite{XLL01} 
the $e^+e^-$ annihilation data reasonably.
We note that two assumptions are used in these calculations: 
First, the spin of the vector meson is taken as the sum 
of the spin of the fragmenting quark and that of the 
anti-quark which combines with the fragmenting quark to form the meson 
and $\alpha$ is a constant.
Second, the influence from the polarization of the fragmenting 
quark on the polarization of the vector mesons which 
are produced in the fragmentation but do not contain 
the fragmenting quark are neglected, thus $\rho_{00}^V(B)$ 
is taken as $1/3$.
It should be emphasized both of them should be tested 
and the situations can be different for longitudinally  
or transversely polarized case. 
In other words, these are the places where the 
differences between $\rho_{00}^V$ in the longitudinally 
polarized case and that in the transversely polarized case can come in. 
Because of the influences of the non-perturbative effects, 
we cannot derive them from QCD at the moment.  
We therefore applied the calculation
method to calculate $\rho_{00}^V$ 
not only in $e^+e^-$ annihilations 
but also in deeply inelastic lepton-nucleon scattering
and high $p_T$ jets in polarized $pp$ collisions\cite{XL02} 
for both longitudinally and transversely polarized cases.
The results can be used to check whether these 
assumptions are true. 
Since we are now mostly interested in the 
qualitative behavior of the $\rho_{00}^V$ 
in unpolarized hadron-hadron collisions, 
the two assumptions seem to be quite safe and 
we will just use the method in the following.

\subsection{Calculating $\rho_{00}^V$ 
in unpolarized hadron-hadron collisions}

According to the method described above, 
to calculate $\rho_{00}^V$ in unpolarized hadron-hadron 
collisions at moderately large transverse momentum, 
we need to find out the number densities for those 
from group (A) and (B) respectively. 
For this purpose, 
we follow the same way as that in Ref.[\ref{LC00J}] 
where single-spin left-right asymmetries $A_N$ 
are studied. 
In Ref.[\ref{LC00J}], the hadrons produced in the 
moderately large transverse momentum region are 
divided into the direct-formation part and 
the non-direct-formation part.   
The number density of the former is denoted by $D(x_F,V)$ 
and that of the latter by $N_0(x_F,V)$. 
(Here, $x_F=2p_{V||}/\sqrt s$, 
$p_{V||}$ is the component of the 
momentum of $V$ parallel to the beam direction,
and $\sqrt{s}$ is the c.m. energy of the
incoming hadron system.)
The direct-formation part denotes
those directly produced and contain the fragmenting valence quarks.
They are described by the ``direct-formation'' or ``direct fusion'' process, 
$q_v+\bar q\to M$, in which the valence quark $q_v$ picks up 
an anti-quark $\bar q$ to form the meson $M$ observed in experiments.
Obviously, this part just corresponds to 
the vector mesons of group (A) mentioned above.
The non-direct-formation part denotes 
the rest which is just the mesons of group (B).  
More precisely, we have, 
$N_V(z,A)\Leftrightarrow D(x_F,V)$, $N_V(z,B)\Leftrightarrow N_0(x_F,V)$.
Hence, the $\rho^V_{00}(x_F)$ in the unpolarized hadron-hadron collisions
is given by,
\begin{equation}
\rho^V_{00}(x_F)=\frac{1}{3}+\Delta \rho^V_{00}(A)\frac{D(x_F,V)}
{D(x_F,V)+N_0(x_F,V)},
\label{eq06}
\end{equation}
The quantization axis is now taken as the normal direction 
of the production plane of the vector meson, and 
$P_q$ is the polarization of the valence quark of the 
projectile with respect to this axis before the hadronization 
or the direct-formation takes place.
We see from Eq.({\ref{eq06}) that, 
$\rho^V_{00}(x_F)$ can be calculated if 
we know $P_{q}$ and the ratio $D(x_F,V)/N_0(x_F,V)$.

\subsubsection{Determination of $P_q$ from the $\vec s_q \cdot \vec n$ 
spin correlation}

The polarization $P_q$ of the quark before the direct formation
of the mesons is determined by the strength of the 
$\vec s_q \cdot \vec n$ type of 
spin-correlation in the process.
Assuming the same origin for the single spin left-right asymmetry
in transversely polarized hadron-hadron collisions and
the spin alignment of vector mesons 
in unpolarized hadron-hadron collisions,
we can extract $P_q$ from the experimental results\cite{ANdata} for $A_N$.
This is exactly the same as what we did in Ref.[\ref{LB97}] and 
now described in detail in the following. 

We recall that\cite{ANdata,LC00J}, in the language commonly used 
in describing $A_N$, the polarization direction 
of the incident proton is called upward, 
and the incident direction is forward.
The single-spin asymmetry $A_N$ is just the difference between
the cross section where $\vec p_V$ points to the left and 
that to the right, which corresponds to 
$\vec s_q \cdot \vec n=1/2$  and $\vec s_q \cdot \vec n=-1/2$  
respectively. 
The data\cite{ANdata} on $A_N$ show that
if a hadron is produced by an upward polarized valence quark of the 
projectile, it has a large probability to 
have a transverse momentum pointing to the left. 
$A_N$ measures the excess of hadrons produced to 
the left over those produced to the right.
The difference of the probability for the hadron to 
go left and that to go right is denoted\cite{LC00J} by $C$. 
$C$ is a constant in the region of $0<C<1$.
It has been shown that\cite{LC00J}, 
to fit the $A_N$ data\cite{ANdata},
$C$ should be taken as, $C=0.6$.

Now, in terms of the spin-correlation discussed above, 
the cross section should be expressed as,
\begin{equation}
\sigma=\sigma_0+(\vec s_q \cdot \vec n )\sigma_1
\label{eqsig}
\end{equation}
where $\sigma_0$ and $\sigma_1$ are independent of $\vec s_q$. 
The second term just denotes the existence of 
the $\vec s_q \cdot \vec n$ type of spin-correlation.
$C$ is just the difference between the cross section
where $\vec s_q\cdot \vec n=1/2$ and that where 
$\vec s_q\cdot \vec n=-{1}/{2}$ divided by the sum of them, 
i.e., $C=\sigma_1/(2\sigma_0)$.

Now, we assume the same strength for the 
spin-correlation in vector meson production
in the same collisions and use Eq.(\ref{eqsig}) to
determine $P_q$.
For a vector meson produced with momentum $\vec p_V$, 
$\vec n$ is given. 
The cross-section that this vector meson is produced 
in the fragmentation of a valence
quark with spin satisfying $\vec s_q \cdot \vec n$=${1}/{2}$ is 
$(\sigma_0+\sigma_1/2)$,
and that with spin satisfying $\vec s_q \cdot \vec n$=$-{1}/{2}$ is
$(\sigma_0-\sigma_1/2)$. 
Hence, the polarization of the valence quarks which lead to 
the production of 
the vector mesons with that $\vec n$ is given by,
\begin{equation}
P_q=\frac{(\sigma_0+\sigma_1/2)-(\sigma_0-\sigma_1/2)}
{(\sigma_0+\sigma_1/2)+(\sigma_0-\sigma_1/2)}
=\frac{\sigma_1}{2\sigma_0}
=C.
\label{eqPq}
\end{equation}

It should be emphasized that, similar to hyperon polarization 
in unpolarized hadron-hadron collisions, 
$P_q\not=0$ just means that the strength of 
the spin-correlation of the form $\vec s_q\cdot \vec n$
is non-zero in the reaction. 
It means that, due to some spin-dependent interactions, 
the quarks which have spins along the same direction as 
the normal of the production plan have a larger probability 
to combine with suitable anti-quarks to form the mesons 
than those which have spins in the opposite direction. 
It does not imply that the quarks in the unpolarized incident hadrons 
were polarized in a given direction, which would contradict 
the general requirement of space rotation invariance. 
In fact, in an unpolarized reaction, 
the normal of the production plane of the mesons 
is uniformly distributed in the transverse directions. 
Hence, averaging over all the normal directions, 
the quarks are unpolarized.   

\subsubsection{Qualitative behavior of 
$\rho_{00}^V$ as a function of $x_F$}

Using the the numerical values $P_q=C=0.6$ and $\alpha \approx 0.5$, 
we obtain $\Delta \rho^V_{00}(A)\approx 0.085$ from Eq.(\ref{eqdelt}).
We see clearly from Eq. (\ref{eq06}) that,
$\rho_{00}^V(x_F)$ is larger than $1/3$ 
as long as $D(x_F,V)$ is nonzero.
Since $0\le D(x_F,V)/[D(x_F,V)+N_0(x_F,V)]\le 1$, 
$\rho_{00}^V(x_F)$ should be in the range of 
$1/3\le \rho_{00}^V(x_F)\le 0.42$.
The detailed form of the $x_F$-dependence of $\rho_{00}^V(x_F)$
is determined by that of the ratio $D(x_F,V)/N_0(x_F,V)$.

To see the qualitative behavior of 
$D(x_F,V)/N_0(x_F,V)$ as a function of $x_F$, 
we recall that 
the valence quarks usually take the large momentum 
fractions of the incident hadrons.  
Hence, at moderately large transverse momenta, 
hadrons which contain the valence quarks of the incident hadrons  
dominate the large $x_F$ region. 
It is therefore clear that, 
to discuss the behavior of $\rho_{00}^V(x_F)$ 
as a function of $x_F$,   
we should divide the vector mesons 
into the following two classes 
according to their flavor compositions: 
(1) those which have a valence quark of
the same flavor as one of the valence quarks of the incident hadron; 
(2) those which have no valence quark 
in common with the incident hadron.
The behavior of $\rho_{00}^V(x_F)$ 
as a function of $x_F$ for these two classes 
of vector mesons should be quite different from each other.

Example of class (1) are: 
\begin{equation}
p+A\to (\rho^+ ,\ \rho^-, {\rm or\ } \rho^0)+X, \\[-0.69cm] 
\end{equation}
\begin{equation}
p+A\to (K^{*+} \ {\rm or\ } K^{*0})+X, \\[-0.69cm] 
\end{equation}
\begin{equation}
K^++A\to (\rho^+ \ {\rm or\ } \rho^0)+X, \\[-0.69cm] 
\end{equation}
\begin{equation}
K^++A\to (K^{*+} \ {\rm or\ } K^{*0})+X. 
\end{equation}
For such vector mesons, 
$D(x_F,V)$ dominate the large $x_F$ region and  
$D(x_F,V)/N_0(x_F,V)$ should increase with increasing $x_F$.
We therefore expect that, for such vector mesons, 
$\rho_{00}^V(x_F)$ increases with increasing $x_F$.
It should start from $1/3$ at $x_F=0$, 
increase monotonically with increasing $x_F$, 
and reach about $0.42$ as $x_F \to 1$.
For different vector mesons in this class,
there will be only some small differences in 
the detailed forms of the $x_F$-dependences,
but the qualitative features remain the same. 

Examples of class (2) are: 
\begin{equation}
p+A\to (K^{*-} \ {\rm or\ } \bar K^{*0})+X,\\[-0.69cm] 
\end{equation}
\begin{equation}
K^++A\to \rho^-+X, \\[-0.69cm] 
\end{equation}
\begin{equation}
K^++A\to (K^{*-} \ {\rm or\ } \bar K^{*0})+X. 
\end{equation}
Here, there is no kinematic region where the vector mesons 
of origin (A) play an important role. 
More precisely, for the vector mesons in this class, 
the mesons of origin (B) dominate for all $x_F$ 
in $0<x_F<1$ at moderately large transverse momenta. 
Hence, we should see no significant spin alignment 
for such vector mesons, 
i.e. $\rho_{00}^V(x_F)\approx 1/3$ for all $x_F$.

We see that although the difference between 
the $\rho_{00}^V$'s of different vector mesons 
in the same class is negligible, 
there is a distinct difference
between the $\rho_{00}^V$ of the vector mesons  
in class (1) and that of the vector mesons in class (2).
This can be checked easily in experiment.
Presently, there are not enough data available 
to check whether these features are true. 
But they seem to be 
in agreement with the existing data\cite{EXCHARM00}. 
Here, we note that $\rho_{00}=0.424\pm0.011\pm0.018$
were obtained for $K^{*+}$ with $p_V>12$GeV 
produced in neutron-carbon collisions 
with $p_{inc}=60$GeV. 
This is an example in the first class and 
the corresponding $x_F$ of the produced $K^{*+}$'s 
in this experiment is quite large so that the $K^{*+}$'s 
are mainly from group (A).
The resulting $\rho_{00}^V$ should be very close to $0.42$ 
which is in agreement with the data. 
Another data is\cite{EXCHARM00},  
$\rho_{00}=0.393\pm0.025\pm0.018$ 
for $K^{*-}$ in neutron-carbon interaction.
This is an example of the second class and it
is also consistent with the theoretical result 
within the experimental errors. 
Further measurements are needed to check whether 
the theoretical predictions are true.

\subsubsection{A rough estimation of 
$\rho_{00}^V$ as a function of $x_F$}

The detailed form of the 
$x_F$-dependence of the ratio 
$D(x_F,V)/N_0(x_F,V)$ is independent of the spin 
properties and can be obtained from the model calculations 
and/or unpolarized experimental data.
A detailed procedure was given in [\ref{LC00J}].
It has been shown that\cite{LC00J}, 
from the general constraints imposed by the conservation laws such as 
energy-momentum and flavor conservation on the direct-formation process,  
$D(x_F,M)$ should be proportional to the product of 
the distribution function of the valence quark and that of the sea anti-quark
which combines the valence quark to form the vector meson. 
The proportional constant is determined by fitting the 
data for unpolarized cross section at large $x_F$.
The non-direct formation part $N_0(x_F,M)$ can be obtained 
by parameterizing the difference between the unpolarized data 
for the number density in the corresponding transverse momentum 
region and the direct formation part $D(x_F,M)$.
We can now follow this procedure to get $D(x_F,V)/N_0(x_F,V)$. 
But unfortunately, there is no appropriate unpolarized data available 
for the differential inclusive cross section of  
vector meson production in the corresponding reactions.
On the other hand, we expect that the qualitative behavior of 
$D(x_F,V)/N_0(x_F,V)$ for the vector mesons should be quite similar to those 
for the pseudo-scalar mesons with the same flavor. 
Hence, we simply use the results for the 
corresponding pseudo-scalar mesons to replace them 
to make a rough estimation of $\rho_{00}^V(x_F)$ 
as a function of $x_F$. 
E.g., for $pp\to K^{*+}X$, 
we take $D(x_F,K^{*+})/N_0(x_F,K^{*+})\approx D(x_F,K^+)/N_0(x_F,K^+)$, 
where the latter has been discussed in detail in the second paper 
of Ref.[\ref{LC00J}].
Using this, we obtain $\rho_{00}^V(x_F)$ for $K^{*+}$ 
as a function of $x_F$ 
from Eq.(\ref{eq06}) as shown in Fig.\ref{fig:rhoV}. 
We see that the obtained $\rho_{00}^{K^{*+}}(x_F)$ 
increases from $1/3$ to 0.42 with increasing $x_F$.
In the fragmentation region, e.g., $x_F$$>0.5$,
it is almost equal to 0.42.
This is consistent with the available data\cite{BCGMNS} 
and can also be checked by future experiments.

We note that it is in principle also possible 
to calculate the transverse momentum dependence 
of the $\rho_{00}^V$. 
For that calculation, we need to know  
transverse momentum dependence of $A_N$ thus 
that of the corresponding $C$. 
Presently, there are not enough data available for such 
a calculation. 
Recent data from BNL E0925 Collaboration\cite{E0925} 
show a significant energy dependence of $A_N$. 
It seems that\cite{DH2003} this energy dependence can 
be obtained from the energy dependence of 
$D(x_F,s|M)$ in the above-mentioned formulation.
Presently, we are working on this and if it is true, 
we can also apply it here to obtain the 
energy dependence of $\rho_{00}^V$. 

\section{Conclusion and discussion}

In summary, 
we argue that the spin alignment of the vector mesons  
in unpolarized hadron-hadron collisions  
and the single-spin left-right asymmetry are 
closely related to each other. 
Both of them are different manifestations  
of the $\vec s_q\cdot\vec n$ type of 
spin correlation in high energy hadron-hadron or
hadron-nucleus collisions. 
We calculate the spin alignment of vector mesons 
in unpolarized hadron-hadron collisions from 
the spin-correlation derived from the single-spin left-right 
asymmetries in hadron-hadron collisions. 
The obtained results are consistent with the available data 
and predictions for future experiments are made. 

\vskip 0.2cm
This work was supported in part by the National Science Foundation
of China (NSFC) under the approval number 10175037 and 
the Education Ministry of China
under Huo Ying-dong Foundation.

\begin{figure}[ht]
\psfig{file=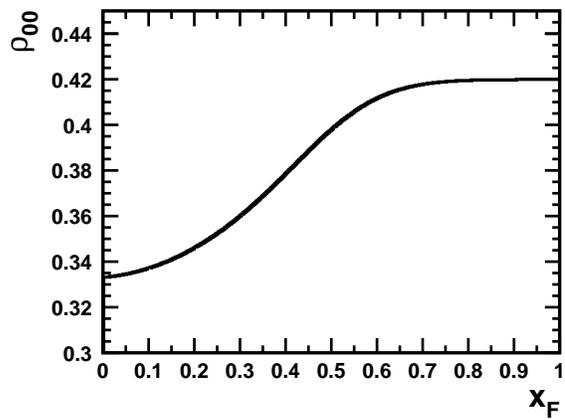,width=10cm}
\caption{Spin alignment of $K^{*+}$ along the normal of the
production plane in unpolarized $pp\to K^{*+}X$ at $p_{inc}=200$ GeV.}
\label{fig:rhoV}
\end{figure}

\newpage
\noindent
Figure Captions

\vskip 0.2cm
\noindent
Fig.1: Spin alignment of $K^{*+}$ along the normal of the
production plane in unpolarized $pp\to K^{*+}X$ at $p_{inc}=200$ GeV.

\end{document}